\documentclass[aps,prd,reprint,superscriptaddress,showpacs,nofootinbib]{revtex4-1}

\usepackage{hyperref}
\usepackage{amsmath}
\usepackage{amssymb}
\usepackage{graphicx}
\usepackage{color}
\usepackage{subfig}
\usepackage{bm}
\usepackage{gensymb}
\usepackage{bigints}
\usepackage[usenames,dvipsnames,svgnames,table]{xcolor}
\usepackage{pifont}
\usepackage{aas_macros}
\usepackage{orcidlink}
\usepackage{physics}
\usepackage{mathrsfs}
\usepackage[graphicx]{realboxes}
\usepackage{multirow}
\usepackage{chemformula}
\usepackage{threeparttable}

\usepackage[labelfont=bf]{caption}

\hypersetup{
    colorlinks=true,
    linkcolor=Blue,
    filecolor=Blue,
    urlcolor=MidnightBlue,
    citecolor=Blue
}

\numberwithin{equation}{section}

\begin{document}


\title{Over-Luminous Type Ia Supernovae and Standard Candle Cosmology}

\author{Abhinandan Ravi\orcidlink{0009-0004-8345-2973}}
\thanks{Corresponding author}
\email[E-mail: ]{abhinandan$_$10@outlook.com}
\affiliation{The Institute of Mathematical Sciences, Chennai 600113, Tamil Nadu, India}
\author{T. R. Govindarajan\orcidlink{0000-0002-8594-0194}}
\email[E-mail: ]{trg@imsc.res.in, govindarajan.thupil@krea.edu.in}
\affiliation{The Institute of Mathematical Sciences, Chennai 600113, Tamil Nadu, India}
\affiliation{Krea University, Sri City 517646, Andhra Pradesh, India}
\author{Surajit Kalita\orcidlink{0000-0002-3818-6037}}
\email[E-mail: ]{skalita@astrouw.edu.pl}
\email{s.kalita@uw.edu.pl}
\affiliation{Astronomical Observatory, University of Warsaw, Al. Ujazdowskie 4, PL-00478 Warszawa, Poland}

\begin{abstract}
Type Ia supernovae (SNe\,Ia) serve as crucial cosmological distance indicators because of their empirical consistency in peak luminosity and characteristic light curve decline rates. These properties facilitate them to be standardized candles for the determination of the Hubble constant ($H_0$) within late-time universe cosmology. Nevertheless, a statistically significant difference persists between $H_0$ values derived from early and late-time measurements, a phenomenon known as the Hubble tension. Furthermore, recent observations have identified a subset of over-luminous SNe\,Ia, characterized by peak luminosities exceeding the nominal range and faster decline rates. These discoveries raise questions regarding the reliability of SNe\,Ia as standard candles for measuring cosmological distances. In this article, we present the Bayesian analysis of 15 over-luminous SNe\,Ia and show that they yield a lower $H_0$ estimate due to the increase in their absolute magnitude. This investigation potentially represents a step toward addressing the Hubble tension.
\end{abstract}

\maketitle

\section{Introduction}
The stability of a main-sequence star is achieved through dynamical equilibrium between the inward gravitational force and the outward radiation pressure generated by nuclear fusion processes within the stellar core. As a star burns out of its nuclear fuel, the radiation pressure diminishes, leading to gravitational contraction. The dynamics of this contraction are significantly influenced by quantum mechanical effects, specifically the onset of electron degeneracy. For main-sequence stars with initial masses less than approximately $10\pm2 M_\odot$ \cite{2018MNRAS.480.1547L}, the electron degeneracy pressure can effectively counterbalance the gravitational pull, preventing further collapse. The resulting stellar remnant, composed of degenerate electron matter, is a white dwarf (WD). The maximum mass that a non-rotating non-magnetized carbon-oxygen WD can hold is approximately $1.4 M_{\odot}$, which is famously known as the Chandrasekhar mass limit \cite{1931ApJ....74...81C, 1935MNRAS..95..207C}.

If a WD accretes matter that exceeds this maximum mass, it becomes dynamically unstable. The temperature at the core of the WD increases to initiate carbon and oxygen fusion, leading to the rapid synthesis of heavier elements. This thermonuclear runaway results in a powerful and luminous explosion, classified as type Ia supernova (SN\,Ia). These explosions typically release approximately $10^{44}\rm\,J$ energy \cite{2023RAA....23h2001L}. They are characterized by their extreme luminosity, rendering them observable over vast cosmological distances. Spectroscopically, SNe\,Ia are identified by the absence of hydrogen and helium emission lines, but exhibit a distinct silicon absorption line at approximately 6150\,\AA. While each individual SN\,Ia exhibits variations in its light curve, the dispersion in their peak luminosities is minimal, attributed to their progenitors reaching near the Chandrasekhar mass limit.

SNe\,Ia, in general, exhibit highly consistent light curves, characterized by predictable temporal variations in their peak luminosities, making them valuable cosmological distance indicators, often referred to as `standard candles.' Distance estimations are carried by comparing the absolute and apparent magnitudes, a concept first proposed in \cite{1998tx19.confE.146P}. A well-established empirical relationship, known as the Phillips relation \cite{1993ApJ...413L.105P}, correlates with a higher peak luminosity with slower luminosity decline rates in the light curve. This inherent property has led to extensive utilization of SNe\,Ia in the determination of the Hubble constant ($H_0$) within the local universe. $H_0$ is a fundamental cosmological parameter, providing insights into the expansion rate and age of the universe. Using several SNe\,Ia data along with measurements from Cepheid variable stars, the SH0ES collaboration \cite{2022ApJ...934L...7R}, obtained $H_{0} = 73.04 \pm 1.04 \rm\, km \, s^{-1} \, Mpc^{-1}$. However, the analysis of cosmic microwave background (CMB) data obtained by the Planck satellite, within the framework of the $\Lambda$ cold dark matter ($\Lambda$CDM) cosmological model, yields $H_{0} = 67.4 \pm 0.5 \rm\, km \, s^{-1} \, Mpc^{-1}$ \cite{2020A&A...641A...6P}. This discrepancy between $H_0$ measurements derived from early- and late-time universe observations is commonly referred to as the Hubble tension. Some comprehensive reviews of various methodologies proposed to resolve this tension can be found in \cite{2021CQGra..38o3001D,2022NewAR..9501659P,2022JHEAp..34...49A,2023Univ....9...94H}.

Over the last couple of decades, observations have revealed a population of over-luminous SNe\,Ia, characterized by exceptionally high peak luminosities and slow decline rate of the light curve than expected \cite{2006Natur.443..308H,2007ApJ...669L..17H,2010ApJ...713.1073S,2012AAS...21924212S,2009ApJ...707L.118Y,2011MNRAS.410..585S,2011MNRAS.412.2735T,2016MNRAS.457.3702P, 2014MNRAS.443.1663C,2016ApJ...823..147C,2022ApJ...927...78D,2020ApJ...900..140H,2021ApJ...922..205A}. The broader light curve (slower decline) leads to over-luminous cases violating the conventional Phillips relation, which relates the decay rate of the light curve to the peak magnitude of the SN\,Ia. This, in turn, makes over-luminous cases not standardizable in the conventional way, and hence, these events were excluded as candidates for standard candles. It was argued that they potentially originate from WDs with masses exceeding the standard Chandrasekhar mass limit \cite{2010PhT....63e..11M}. Several theoretical mechanisms have been proposed to explain the formation of these super-Chandrasekhar mass WDs. These mechanisms include the presence of strong magnetic fields exceeding the Schwinger limit, which enhance the magnetic pressure, thereby allowing for increased mass accumulation \cite{2013PhRvL.110g1102D, 2013IJMPD..2242004D}. Another hypothesis involves high angular momentum, where rapid spin generates a centrifugal force that expands the WD, enabling further mass accretion while maintaining hydrostatic equilibrium \cite{2018A&A...618A.124F}. Additionally, modified gravity theories, which effectively alter the Poisson equation, have been proposed as a means of producing WDs with masses significantly exceeding $1.4 M_{\odot}$ \cite{2018JCAP...09..007K,2021ApJ...909...65K,2022PhLB..82736942K}. Furthermore, the effects of noncommutative geometry, which become significant at length scales comparable to the electron Compton wavelength, can modify the equation of state of degenerate electrons, potentially allowing for greater mass accumulation \cite{2021IJMPD..3050034K,2021IJMPD..3050101K}. Notably, super-Chandrasekhar mass WDs have not been directly observed in surveys such as Gaia or Kepler, likely because of their expected low luminosities.

The existence of over-luminous SNe\,Ia raises questions regarding the reliability of SNe\,Ia as standard candles given their exceptionally high luminosities. In other words, the existence of over-luminous SNe\,Ia raises questions about the reliability of the empirical Phillips relation and its utilization to standardize regular SNe\,Ia as it cannot be extended without modification to these peculiar over-luminous cases. This implies a gap in our theoretical understanding of this empirical relation is incomplete, and further studies are required to extend it to all SNe\,Ia. Moreover, these events exhibit slower light curve decay rates compared to the standard SNe\,Ia, deviating from the standard Phillips relation \cite{2017hsn..book..317T}. As previously mentioned, the theoretical understanding of the explosion mechanisms of these over-luminous events remains incomplete, and there is no definitive observational evidence to accurately determine the progenitor mechanism. In this study, we utilize a sample of 15 over-luminous SNe\,Ia to calculate $H_0$. A modification of the empirical relation estimating the absolute magnitudes of SNe\,Ia is proposed to predict the absolute magnitudes of their over-luminous counterparts using properties of their light curves. This value is used to perform a Bayesian analysis and derive $H_0$. We demonstrate that the inclusion of these over-luminous SNe\,Ia in the analysis yields a lower value for $H_0$, aligning more closely with measurements from early-universe observations.

In general, existence of over-luminous SNe\,Ia leads to distances to these events being underestimated. To correct this, one needs to reduce the value of $H_0$ from that obtained from the local measurement, making it more agreeable to the value obtained from the CMB. This article is structured as follows. In Section \ref{Sec2}, we review the fundamental properties of SNe\,Ia at peak luminosity and examine the dependence of their peak brightness on the synthesized nickel mass. Section \ref{Sec3} presents our dataset of over-luminous SNe\,Ia and details the methodology employed to estimate $H_0$. Additionally, we perform a Bayesian analysis incorporating various priors to infer $H_0$ in a cosmology-independent manner. Finally, in Section \ref{Sec4}, we discuss our findings and provide concluding remarks.


\section{Revisiting luminosity of type Ia supernova}\label{Sec2}

SNe\,Ia are used as one of the standard candles in astronomy to measure luminosity distances due to their predicting behavior in peak brightness and declining rates. If $L_\text{SN\,Ia}$ is the luminosity of the SN\,Ia, $F$ is its flux measured on the Earth, and $d_\text{L}$ is its luminosity distance, they are related as
\begin{equation}
    F = \frac{L_\text{SN\,Ia}}{4\pi d_\text{L}^{2}}.
\end{equation}
Here $d$ can be estimated using the distance modulus formula given by \cite{2010asph.book.....C}
\begin{equation}\label{Eq: distance modulus}
    \mu = \mathsf{m} - \mathsf{M} = 5\log_{10} \left( \frac{d_\text{L}}{10\rm\,pc} \right),
\end{equation}
where $\mathsf{m}$ is the apparent magnitude and $\mathsf{M}$ is absolute magnitude of the SN\,Ia. For an object at a redshift $z$, $d_\text{L}$ is given by
\begin{align}
    d_\text{L} = (1+z)\frac{c}{H_0}\int_0^z\frac{\dd{z}}{E(z)},
\end{align}
where $c$ is the speed of light and under $\Lambda$CDM framework, $E(z)=\sqrt{\Omega_\text{m}(1+z)^3+\Omega_\Lambda}$ with $\Omega_\text{m}$ and $\Omega_\Lambda$ being respectively the matter and cosmological density parameters satisfying $\Omega_\text{m}+\Omega_\Lambda=1$. The peak luminosity of a SN\,Ia mostly depends on the amount of $\ch{^{56}Ni}$ produced, whereas the afterglow is largely due to the radioactive decay of nickel to Cobalt to Iron, as follows:
\begin{equation}
    \ch{^{56}Ni} \rightarrow \ch{^{56}Co} \rightarrow \ch{^{56}Fe}.
\end{equation}
The relation between the peak luminosity and the mass of $\ch{^{56}Ni}$ produced during the process is given by \cite{2007ApJ...662..487W}
\begin{equation}
    L_\text{max} \sim f M_\text{Ni} \exp\left(-\frac{t_{p}}{t_\text{Ni}}\right),
\end{equation}
where $f$ is the percentage of gamma-ray decay energy trapped at the bolometric peak (typically $f\approx1$), $t_{p}$ is the rise time to peak luminosity, and $t_\text{Ni}$ is the decay time of $\ch{^{56}Ni}$ \cite{2007ApJ...662..487W}. The decay timescale of luminosity of a SN\,Ia is typically given by \cite{2007ApJ...662..487W}
\begin{equation}
    t_{d} \sim \Phi_\text{Ni} \kappa^{1/2}M_\text{Ni}^{3/4}E_{k}^{-1/4},
\end{equation}
where $\Phi_\text{Ni}$ describes the fractional distance between the bulk of $\ch{^{56}Ni}$ and the ejecta surface, $\kappa$ is the effective opacity per unit mass, and $E_{k}$ is the kinetic energy of ejecta. Moreover, using another model with homogeneous expansion of spherical shock proposed by Arnett \cite{1982ApJ...253..785A}, the relation between peak luminosity and nickel mass is given by \cite{2009ApJ...707L.118Y}
\begin{equation}\label{Eq: peak luminosity}
    L_\text{max} = \left(6.45 e^{\frac{-t_{r}}{8.8\,\mathrm{d}}} + 1.45 e^{\frac{-t_{r}}{111.3\,\mathrm{d}}}\right) \times 10^{43} \frac{M_\text{Ni}}{M_{\odot}} \rm\, erg\, s^{-1},
\end{equation}
where $t_r$ is the rising time of bolometric luminosity in days. Note that this $L_\text{max}$ is a simplified estimate based on that decay of nickel is the primary source of energy. There can be many more complex phenomena occurring during the SN\,Ia process, and to discuss them, one needs to suitably incorporate nuclear physics, which is beyond the scope of this study.

The peak absolute magnitude is related to the peak luminosity of Equation~\eqref{Eq: peak luminosity}. Thus plugging it into the luminosity--absolute magnitude relation, we obtain \cite{1996ima..book.....C}
\begin{equation}
    \mathsf{M}_B = \mathsf{M}_{\odot} - 2.5\log\left( \frac{L_\text{SN\,Ia}}{L_{\odot}} \right).
\end{equation}
Moreover, using Equation~\eqref{Eq: peak luminosity}, the amount of nickel produced during the event can be estimated considering the radioactive decay of nickel is the primary source driving the bolometric light curve. Although this holds true for the standard luminosity SNe\,Ia, new elements are thought to be synthesized in the over-luminous SNe\,Ia as the masses of the progenitors exceed the Chandrasekhar limit. It is expected that the luminosity of these over-luminous events may be affected by the other elements synthesized during the explosion. The exact relation between the luminosity and the mass of the progenitor is still unknown and hence the luminosity might not be captured completely by the radioactive decay of nickel alone.

Another important parameter is absolute magnitude $\mathsf{M}_B$, which for standard-luminosity SNe\,Ia can be estimated using parameters from their light curves. Tripp \cite{1998A&A...331..815T} found an empirical linear relation between $\mathsf{M}_B$ and the 15 days decrease of apparent magnitude $\Delta m_{15}$, given by
\begin{equation}\label{Eq: N-SNe}
    \mathsf{M}_{B} = -19.48 + b(\Delta m_{15} - 1.05) + R(B-V).
\end{equation}
where the parameter $b$ is the slope of the luminosity decline rate and $R(B-V)$ is the correction for reddening. The value $-19.48$ corresponds to the average absolute magnitude of seven original Cepheid-calibrated SN\,Ia and $1.05$ is their average decline rate. The best fit values for $b$ was found $b\approx0.52$ \cite{1998A&A...331..815T,2010AJ....139..120F}. Using the same values of these parameters in the case of over-luminous SNe\,Ia leads might not be correct as they are intrinsically brighter and slower in declining rate. Hence, we write the above effective equation for over-luminous cases as
\begin{equation}\label{Eq: O-SNe}
    \mathsf{M}_{B} = a + b(\Delta m_{15} - c) + R(B-V).
\end{equation}
Since over-luminous SNe\,Ia are brighter, it is expected $a < -19.48$. Moreover, as these over-luminous events display a slower decay rate, it implies $b > 0.52$ and $c < 1.05$ to obtain the best estimate of $\mathsf{M}_{B}$ for over-luminous SNe\,Ia. Calculating the $R(B-V)$ term for over-luminous SNe\,Ia becomes a lot more complex compared to normal-luminosity SNe\,Ia because we cannot be sure about the intrinsic color of over-luminous events. This makes it hard to know whether the redness observed is due to dust or is an intrinsic feature of these events.


\section{Hubble constant estimation using over-luminous type Ia supernovae}\label{Sec3}

In this study, we analyze 15 over-luminous SNe\,Ia, whose progenitors are thought to be super-Chandrasekhar WDs, and thereby estimate $H_0$. The relevant data for these 15 supernovae are presented in Table~\ref{tab: supernovae}. Here $z$ represents the redshift, $\mathsf{m}_{B}$ is the apparent bolometric magnitude, and $\mathsf{M}_{B}$ is the absolute bolometric magnitude at the peak of the SNe\,Ia. $M_\mathrm{WD}$ is the mass of the progenitor WD producing these SNe\,Ia inferred from the measured nickel mass and ejecta velocity. Note that to produce the measured value of the high nickel mass during the supernova explosion from a WD, its mass must exceed the Chandrasekhar mass limit.

\begin{table}[htpb]
    \centering
    \caption{Data of over-luminous SNe\,Ia along with the estimated values of nickel mass and progenitor mass.}
    \label{tab: supernovae}
    \begin{tabular}{|l|l|l|l|l|l|l|l|}
    \hline
    Name & $z$ & $\Delta m_{15}$ & $\mathsf{m}_{B}$ & $R(B-V)$ & Ref.\\
    \hline
    SN\,2003fg & 0.2440 & 0.94 & 20.35 $\pm$ 2.04* & 0 & \cite{2006Natur.443..308H} \\ 
    SN\,2006gz & 0.0237 & 0.69 & 16.06 $\pm$ 1.61* & 1.02 & \cite{2007ApJ...669L..17H} \\ 
    SN\,2007if & 0.0742 & 0.71 & 17.34 $\pm$ 0.04 & 0.06 & \cite{2010ApJ...713.1073S, 2012AAS...21924212S} \\ 
    SN\,2009dc & 0.0214 & 0.65 & 15.19 $\pm$ 0.16 & 0.46 & \cite{2009ApJ...707L.118Y,2011MNRAS.410..585S,2011MNRAS.412.2735T} \\
    SN\,2012dn & 0.010187 & 0.92 & 14.38 $\pm$ 0.02 & 0.75 & \cite{2016MNRAS.457.3702P, 2014MNRAS.443.1663C} \\
    SN\,2013cv & 0.035 & 1.03 & 16.28 $\pm$ 0.03 & 0.18 & \cite{2016ApJ...823..147C} \\
    SN\,2020esm & 0.03619 & 0.92 & 16.16 $\pm$ 0.03 & 0.09 & \cite{2022ApJ...927...78D} \\
    LSQ\,14fmg & 0.0649 & 1.062 & 17.39 $\pm$ 0.01 & 0.15 & \cite{2020ApJ...900..140H} \\
    ASASSN-15pz & 0.0148 & 0.67 & $14.18 \pm 0.03$ & -0.02 & \cite{2021ApJ...922..205A} \\
    LSQ\,12gpw & 0.0506 & 0.70 & $17.35 \pm 0.01$ & 0.01 & \cite{2021ApJ...922..205A} \\
    SN2013ao & 0.0435 & 0.99 & $16.87 \pm 0.01$ & 0.10 & \cite{2021ApJ...922..205A} \\
    CSS140126 & 0.0772 & 0.73 & $18.23 \pm 0.01$ & -0.06 & \cite{2021ApJ...922..205A} \\
    CSS140501 & 0.0797 & 1.05 & $18.09 \pm 0.04$ & 0.03 & \cite{2021ApJ...922..205A} \\
    SN2015M & 0.0231 & 0.82 & $15.54 \pm 0.03$ & 0.14 & \cite{2021ApJ...922..205A} \\
    ASASSN-15hy & 0.0185 & 0.73 & $15.19 \pm 0.01$ & 0.19 & \cite{2021ApJ...922..205A} \\
    \hline
    \end{tabular}
    \begin{tablenotes}\centering
        \item *Errors are taken to be 10\% of the original value as their exact values are not reported.
        \item **SN\,2004gu is an over-luminous SN\,Ia as reported in \cite{2017arXiv170702543S}. However, it was not included in the analysis due to the limited availability of observational data of the event.
    \end{tablenotes}
\end{table}


\subsection{Bayesian analysis to estimate the Hubble constant}

\begin{figure}[htpb]
    \centering
    \includegraphics[scale=0.5]{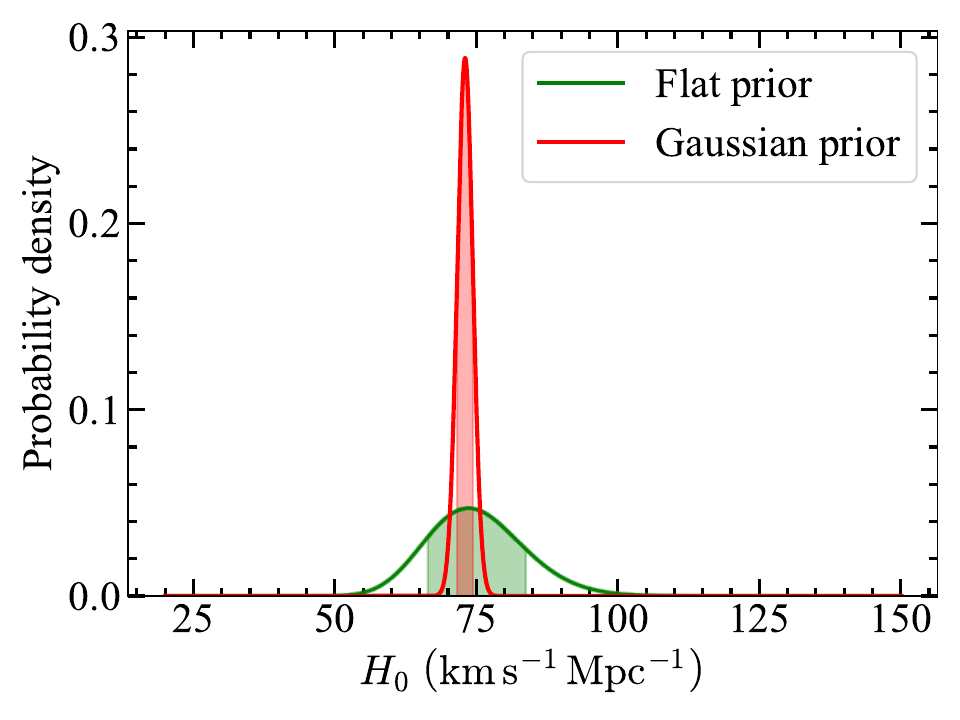}
    \caption{Probability distributions of likelihood function with respect to $H_0$ along with the corresponding 1$\sigma$ confidence intervals for $a=-19.48$, $b=0.52$, and $c=1.05$. The green curve represents the case of a flat prior on $H_0$ that is maximized at $H_{0} = 73.60_{-7.18}^{+10.14} \rm \, km \, s^{-1} \, Mpc^{-1}$. Red curve represents the case for Gaussian prior on $H_0$ from Cepheid variable stars and maxima of the likelihood function shifts to $H_0 = 73.02_{-1.40}^{+1.38} \rm\,km\,s^{-1}\,Mpc^{-1}$.}
    \label{Fig: pdfunc}
\end{figure}

Assuming Gaussian scatter of in $\mathsf{M}_B$ values, we now define the probability distribution for each SN\,Ia as
\begin{equation}\label{Eq: prob}
    P_{i}(\mathsf{m}_B\mid H_0) =  \frac{1}{\sqrt{2\pi \sigma_{i}^{2}}} \exp\left( \frac{\mathsf{M}_{B_{i}} - \mathsf{M}_{T_{i}} }{\sigma_{i}} \right)^{2},
\end{equation}
where $\sigma_{i}$ is the error bar for $\mathsf{M}_B$ which is estimated using Eq.~\eqref{Eq: O-SNe} and $\mathsf{M}_{T}$ is the expected absolute magnitude calculated from Eq.~\eqref{Eq: distance modulus}. As we are unsure of the values of $(a,b,c)$, we assume $\sigma_{i} = 5\%\mathsf{M}_{B_{i}}$. Assuming a flat prior on $H_0$, the joint likelihood function can be defined as the product of each of the aforementioned individual likelihood function, given by
\begin{align}\label{Eq: Jointprob}
    \mathcal{L} &= \prod_{i = 1}^{N} P_{i}(\mathsf{M}_B\mid H_0),
\end{align}
where $N$ is the total number of SNe\,Ia in the data sample. The green curve in Fig.~\ref{Fig: pdfunc} shows the joint likelihood function plotted against the different values of $H_{0}$ with the original values of $(a,b,c)$. It is evident that the likelihood is maximized at $H_{0} = 73.60_{-7.18}^{+10.14} \rm \, km \, s^{-1} \, Mpc^{-1}$ within 1$\sigma$ uncertainty.

We further consider a Gaussian prior on $H_0$ as
\begin{equation}
    \mathcal{P}(H_0) = \frac{1}{\sqrt{2\pi\sigma^{2}}} \exp{-\frac{(H_{0} - \bar{\mu})^{2}}{2 \sigma^{2}}}.
\end{equation}
We choose the prior obtained from the Cepheid variable stars in this case. Cepheid variable stars are pulsating stars whose luminosity changes with well-defined periodicity, resulting from radial oscillations in their outer envelopes. The key characteristic that makes Cepheids valuable for distance measurements is the period-luminosity relationship \cite{1912HarCi.173....1L}. This empirical relation establishes that longer-period Cepheids are intrinsically more luminous. By measuring the pulsation period and apparent brightness, their absolute magnitudes and distances can be determined. Therefore, Cepheids serve as primary calibrators for SNe\,Ia \cite{2016ApJ...826...56R,2019ApJ...882...34F}. We consider the mean $\bar{\mu} = 73.0\rm\,km\,s^{-1}\,Mpc^{-1}$ and standard deviation $\sigma= 1.4\rm\,km\,s^{-1}\,Mpc^{-1}$ following \cite{2021ApJ...919...16F} for Cepheid variable stars. Therefore, the joint likelihood function can be defined as 
\begin{equation}\label{Eq: Jointprob2}
    \mathcal{L} = \left[\prod_{i = 1}^{8} \frac{1}{\sqrt{2\pi \sigma_{i}^{2}}} \exp\left( \frac{\mathsf{M}_{B_{i}} - \mathsf{M}_{T_{i}} }{\sigma_{i}} \right)^{2}\right] \mathcal{P}(H_0). 
\end{equation}
The red curve in Fig.~\ref{Fig: pdfunc} shows variation of the modified likelihood function with a prior knowledge on $H_0$ from the Cepheid variable stars, which is maximized at $H_{0} = 73.02_{-1.40}^{+1.38} \rm \, km \, s^{-1} \, Mpc^{-1}$ within 1$\sigma$ uncertainty. 

\begin{figure}[htpb]
    \centering
    \includegraphics[scale=0.5]{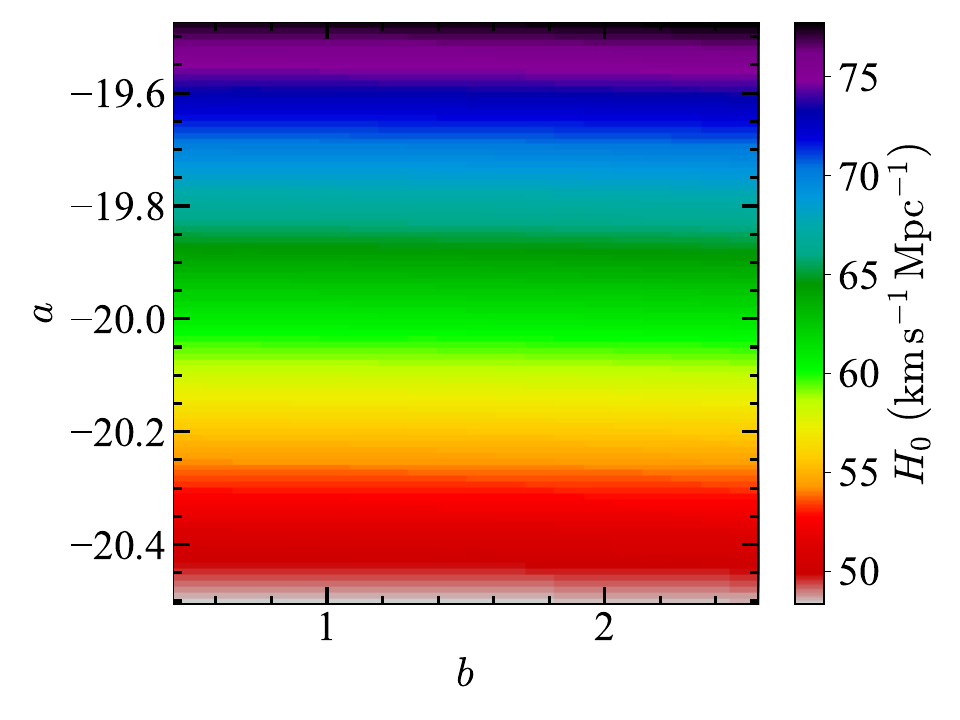}
    \caption{Maximized values of $H_0$ for different combinations of $a$ and $b$ in the case of a flat prior on $H_0$ for a fixed $c=0.84$.}
    \label{Fig: Contour}
\end{figure}

As mentioned previously, in the case over-luminous events, it expected to have $a < -19.48$, $b > 0.52$, and $c < 1.05$. Out of these three parameters, $c$ represents the average of $\Delta m_{15}$, for the sample of over-luminous SNe\,Ia listed in Table~\ref{tab: supernovae}, we have $c=0.84$. The other quantities are difficult to find for these events as they are yet to be standardized from other astronomical objects like Cepheids found in the same galaxy. As a result, we keep these two parameters free and estimate $H_0$ for different combinations of $a$ and $b$, which is shown in Fig.~\ref{Fig: Contour}. It is evident that change of $b$ has negligible effect on $H_0$ values. Moreover, $H_0$ decreases with the decrease in $a$. In case the over-luminous events have absolute magnitude has at least one order of magnitude less than the standard one, as previously shown in \cite{2017hsn..book..317T}, $H_0$ can reduces significantly.

\section{Discussion}\label{Sec4}

This article discusses over-luminous SNe\,Ia and their potential use in understanding and resolving the Hubble tension. This study analyzes 15 over-luminous SNe\,Ia listed in Table~\ref{tab: supernovae} to compare $\mathsf{M}_B$ calculated using the Tripp relation with the same estimated using the distance modulus formula. As the latter dependent on $H_0$, the goal is to understand the difference in the $\mathsf{M}_B$ values and determine $H_{0}$ for which the difference is minimized. It is important to note that the $\mathsf{M}_B-\Delta m_{15}$ relation is quite empirical and the observed deviation underscores the need to examine the empirical Phillips relation to explicitly account for the over-luminous SNe\,Ia. More research needs to be carried out to understand how this relation arises from theoretical considerations to verify its extension for the over-luminous case. Utilizing a Bayesian statistical framework and a sample of 15 over-luminous SNe\,Ia alone, we demonstrate that a decrease of an order-of-magnitude in absolute magnitude, $H_0$ can reduce significantly. However, keeping the same values of $(a,b,c)$ in Eq.~\eqref{Eq: O-SNe} as the case for standard SNe\,Ia for $\mathsf{M}_B$, there is a no significant deviation in $H_0$ estimate using over-luminous SNe\,Ia alone with respect to the value reported by SH0ES collaboration within 1$\sigma$ uncertainty.

This analysis highlights a novel discrepancy: the noticeable divergence in maximized $H_0$ values obtained from over-luminous SNe\,Ia and those derived from standard SNe\,Ia. This raises fundamental questions regarding the intrinsic standardization of SNe\,Ia luminosities, particularly considering the inclusion of over-luminous events. The unavailability of standardization techniques in the over-luminous cases presents a problem as the techniques cannot be yet generalized to all SNe\,Ia. It is necessary to understand the theoretical underpinnings of the Phillips relation and extend it to over-luminous cases. The application of over-luminous SNe\,Ia as distance indicators may lead to a systematic underestimation of cosmological distances, contributing to their exclusion from standard candle applications. Furthermore, over-luminous SNe\,Ia exhibit substantial variability in peak luminosity, spectral characteristics, and light curve morphology. The potential existence of super-Chandrasekhar mass WD progenitors necessitates a rigorous re-evaluation of standardization methodologies for over-luminous SNe\,Ia, analogous to those applied to standard events. A comprehensive theoretical understanding of the explosion mechanisms associated with super-Chandrasekhar mass WDs, as suggested in \cite{2013PhRvL.110g1102D}, may facilitate the development of over-luminous SNe\,Ia as a distinct class of standardizable candles. A few attempts have been made to explain the WD to SN\,Ia mechanism in the single degenerate (WD accreting matter from a second star) case in \cite{2009ApJ...702..686C, 2012ApJ...744...69H}. A primary limitation of this study is the small sample size, which was constrained by the lack of available data. Future investigations employing an expanded catalog of over-luminous SNe\,Ia have the potential to refine the $H_0$ estimate and contribute substantially to the resolution of Hubble tension.

\section*{Acknowledgments}
T.R.G. gratefully acknowledges Dr. Ravi Sheth from the University of Pennsylvania and Dr. S. Kalyana Rama from IMSc for insightful discussions and valuable comments. S.K. acknowledges funding by the National Center for Science, Poland, grant no. 2023/49/B/ST9/02777. A.R. and S.K. would like to thank IMSc, Chennai for providing support and resources during the course of this study.



\bibliographystyle{apsrev4-1}
\bibliography{biblio}

\end{document}